\def\km{\,{\rm km}}

\def\kpc{\,{\rm kpc}}

\def\Mpc{\,{\rm Mpc}}

\def\s{\,{\rm s}}

\def\Msol{\,{{\rm M}_\odot}}

\def\aua#1#2{{ #1, }{A\&A,}{ #2}}

\def\apj#1#2{{#1, }{ApJ,} { #2}}
\def\apjs#1#2{{#1, }{ApJS,} { #2}}

\def\mnras#1#2{{#1, }{MNRAS,} { #2}}

\def\Atoday{\ifcase\month\or
  January\or February\or March\or April\or May\or June\or
  July\or August\or September\or October\or November\or December\fi
  \space\number\day, \number\year}
\def\Etoday{\number\day\space\ifcase\month\or
  January\or February\or March\or April\or May\or June\or
  July\or August\or September\or October\or November\or December\fi
  \space\number\year}

\immediate\write16{... Special symbols are defined}
\def\la{\mathrel{\hbox{\rlap{\hbox{\lower4pt\hbox{$\sim$}}}\hbox{$<$}}}}
\def\ga{\mathrel{\hbox{\rlap{\hbox{\lower4pt\hbox{$\sim$}}}\hbox{$>$}}}}
\def\lse{\mathrel{\hbox{\rlap{\hbox{\raise4pt\hbox{$\<$}}}\hbox{$\simeq$}}}}
\def\gse{\mathrel{\hbox{\rlap{\hbox{\raise4pt\hbox{$\>$}}}\hbox{$\simeq$}}}}
\def\loa{\mathrel{\hbox{\rlap{\hbox{\lower4pt\hbox{$\approx$}}}\hbox{$<$}}}}
\def\goa{\mathrel{\hbox{\rlap{\hbox{\lower4pt\hbox{$\approx$}}}\hbox{$>$}}}}

\def\ed{\end{document}}

\def\beq#1{\begin{equation}\label{#1}}
\def\eeq{\end{equation}}
\def\beqa#1{\begin{eqnarray}\label{#1}}
\def\eeqa{\end{eqnarray}}

\def\bfig{\begin{figure}[h] \centerline{\hbox{}}\vfill}
\def\efig{\end{figure}\vfill\newpage}

\def\spose#1{\hbox to 0pt{#1\hss}}
\def\simlt{\mathrel{\spose{\lower 3pt\hbox{$\mathchar"218$}}
     \raise 2.0pt\hbox{$\mathchar"13C$}}}
\def\simgt{\mathrel{\spose{\lower 3pt\hbox{$\mathchar"218$}}
     \raise 2.0pt\hbox{$\mathchar"13E$}}}
\def\simpropto{\mathrel{\spose{\lower 3pt\hbox{$\mathchar"218$}}
     \raise 2.0pt\hbox{$\propto$}}}

\documentstyle[psfig]{lamuphys}
\makeatletter
\let\chapter\hid@chapter
\makeatother
\begin{document}
\pagenumbering{arabic}
\title{Metal absorption from galaxies in the process of formation}

\author{Martin G.\,Haehnelt\inst{}}

\institute{Max-Planck-Institut f\"ur Astrophysik,
Karl-Schwarzschild-Stra\ss e 1, 85740 Garching, Germany}

\maketitle

\begin{abstract}
In a hierarchical cosmogony  present-day galaxies build up by 
continuous merging of  smaller structures.   At a redshift of three 
the matter content of a typical present-day galaxy is split into about 
ten individual protogalactic clumps. Numerical simulations show that 
these protogalactic clumps have a typical distance of about 
100 kpc, are  embedded in a sheet-like structure and are often aligned 
along filaments. Artificial QSO spectra were generated from  
hydrodynamical simulations of such regions of ongoing galaxy formation. 
The metal and hydrogen absorption features in 
the artificial  spectra closely resemble observed  systems 
over a wide range in HI column density. 
Detailed predictions of the column density as a function of 
impact parameter to protogalactic clumps are  presented  for 
HI,CII,CIV,SiIV,NV and OVI. The expected correlations between 
column densities of different species and their role in understanding the 
physical properties of the gas from which galaxies  form are discussed. 
The model is able to explain both high-ionization multi-component 
heavy-element absorbers and damped Lyman alpha systems as  groups 
of small protogalactic clumps. 
\end{abstract}
\section{Introduction}

While at low redshift metal absorption systems have been convincingly 
demonstrated to arise in the haloes of rather normal galaxies
(Boisse \& Bergeron 1991, Steidel 1995),
much less is known about the nature of metal absorption   system
at redshifts $z \ga 2$ (Sargent, Boksenberg \& Steidel 1988;
Petitjean \& Bergeron 1994; Aragon-Salamanca et al. 1994). 
Recently,  it was shown
that the prominent complex CIV absorption features observed at high
redshift  can be well  reproduced by the absorping properties of
regions in which galaxies form by hierarchical merging 
(Haehnelt,  Steinmetz \& Rauch 1996). Here we present further
results concerning the metal absorption properties  of 
such regions.

\section{Numerical Simulations}

The simulations were performed using GRAPESPH (Steinmetz 1996).
The cosmological background model is a $\Omega=1$,
$H_0=50\,\km\s^{-1}\,\Mpc^{-1}$ cold dark matter (CDM) cosmogony with a
normalization of $\sigma_{8}=0.63$. The baryon fraction is $\Omega_b=0.05$. 
The gas particle masses is $5\times 10^6\,\Msol$. The high resolution
region of the simulation box is about  5.5 comoving Mpc across and 
contains three galaxies with circular velocities between $100$ and 
$200\km\s^{-1}$ at redshift zero.
For the UV background  a power-law spectrum with spectral index $\alpha=-1.5$,
a normalization of $J_{21}($z= 3$) = 0.3$ and a redshift dependence  
as given by Vedel, Hellsten \& Sommer--Larsen (1994) was assumed.  
The photoionization code CLOUDY (Ferland 1993) was used to calculate the
ionization state of the gas (see  the contribution by Steinmetz 
in this volume for a more detailed description of the simulation).

\section{The gas distribution}

There are about 20 collapsed protogalactic clumps (PGC's) aligned in  
a filamentary matrix which itself is embedded in a sheet-like 
structure (Figure 1a).
\begin{figure}[t]
\centerline{
\psfig{file=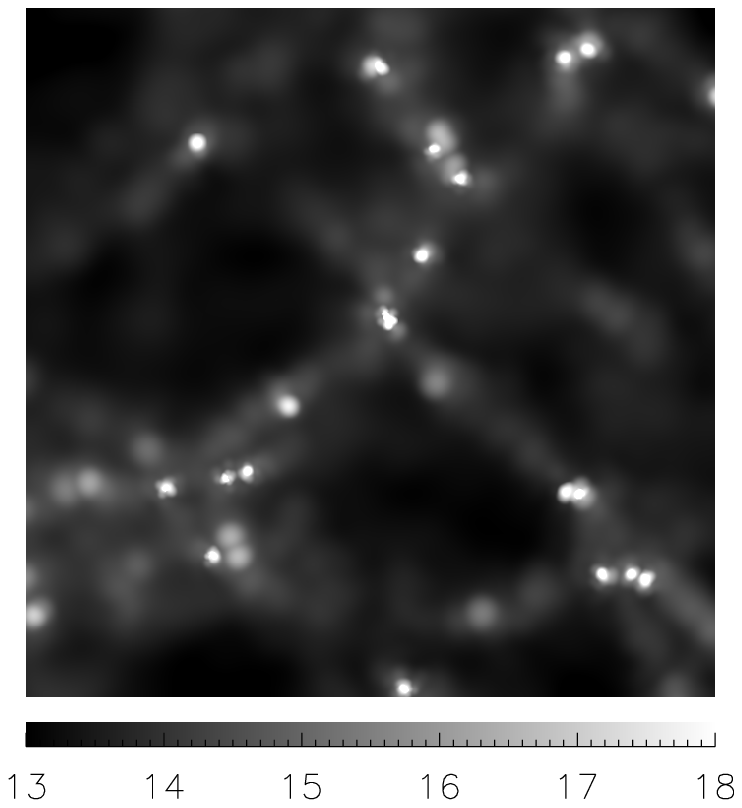,width=6.0cm,angle=0.}
\hspace{-3.5cm}
\psfig{file=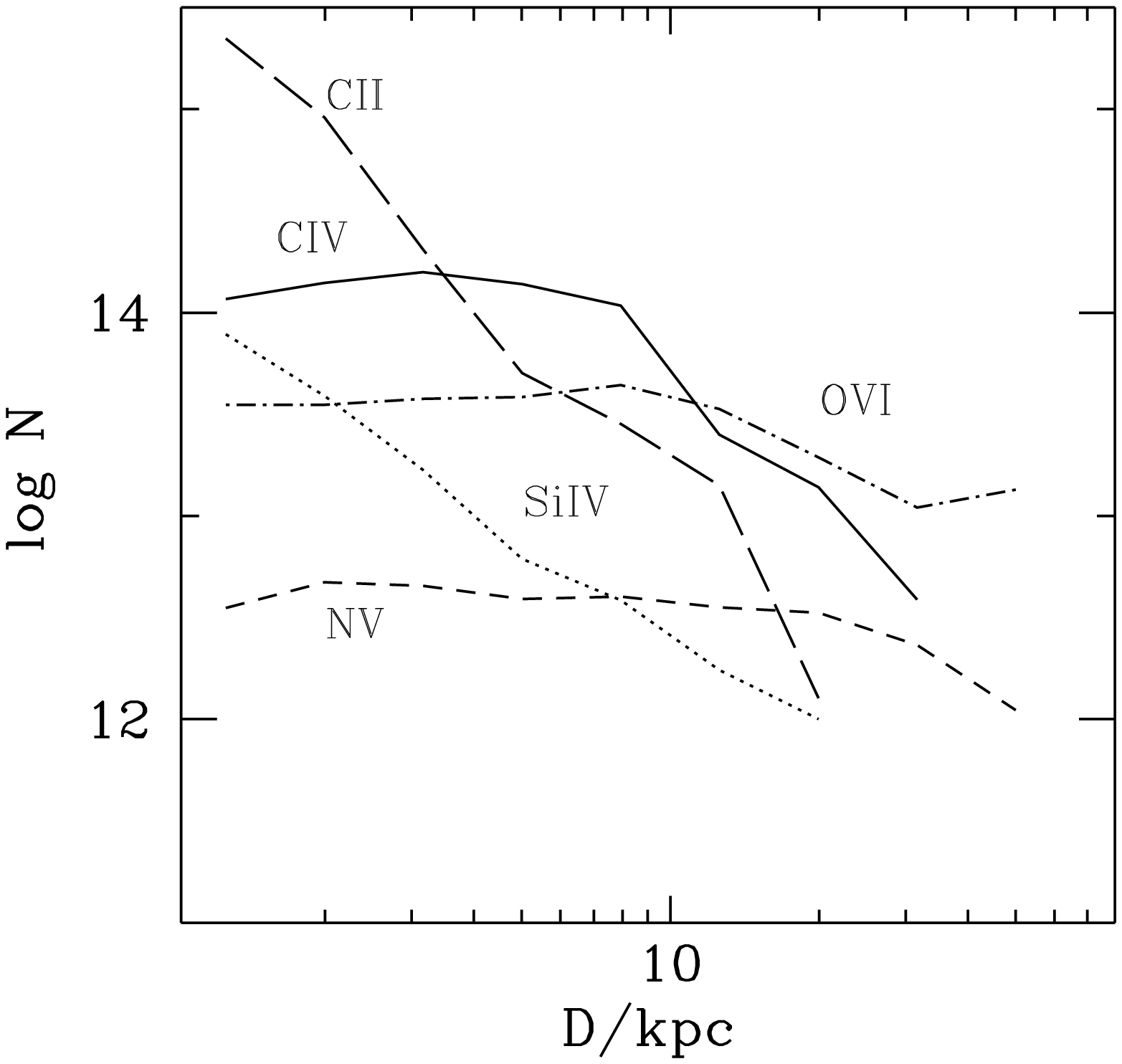,width=10.0cm,angle=0.}
}
\caption{The left panel shows the HI column density (log N(HI))
distribution of the inner $700 \kpc$ (proper length) 
of the simulation box projected along one of the axis
at $z= 3.07$. In the right panel the column densities of a set of ions 
is plotted as a function of the impact parameter to a typical protogalactic
clump in the simulation. A homogeneous metallicity of 0.01 solar is assumed.}
\end{figure}
Figure 2a shows the observed column density distribution of HI and a set 
of other ionic species  observable in absorption
systems of intermediate HI column density at high redshift. 
The crosses show the  observed HI distribution 
($f(N) N$) obtained by Petitjean et al. (1993). Apart from the high
column density end the shape of the observed and simulated HI
distribution  correspond rather well. There  
self-shielding becomes important which is not taken into account 
in the simulation. The normalization was freely adjusted 
as the simulation box is too small to be a fair sample of the universe. 
In Figure 2b the typical column density (mean log N) is shown 
as function of  absorption-weighted overdensity 
$\delta = |\rho - \bar \rho|/\bar \rho$. There is a tight correlation 
between column density and density confirming the visual impression
from Figure 1a that  higher column densities probe the centre 
of the collapsed PGC's.  Low column
densities arise from the  more diffuse gas in sheets and filaments 
between the collpased objects. Figure 1b and 2b show that the 
spatial distribution  of different ionic species differs.  
Species like CII and SiIV are only strong in the  dense
inner regions of the PGC's and the column density 
falls off rather rapidly on scales of $10 \kpc$ or less. Higher 
ionization species probe the gas further away from the PGC's. The 
high CIV column density extends to  scales of typically 30 kpc. A 
closer inspection of the simulation shows that CIV is a good tracer 
of the filamentary matrix connecting the PGC's.
NV and OVI probe even lower densities. Especially OVI is a good
tracer of the diffuse intergalactic medium in the sheets between 
the PGC's.

\begin{figure}[t]
\centerline{
\psfig{file=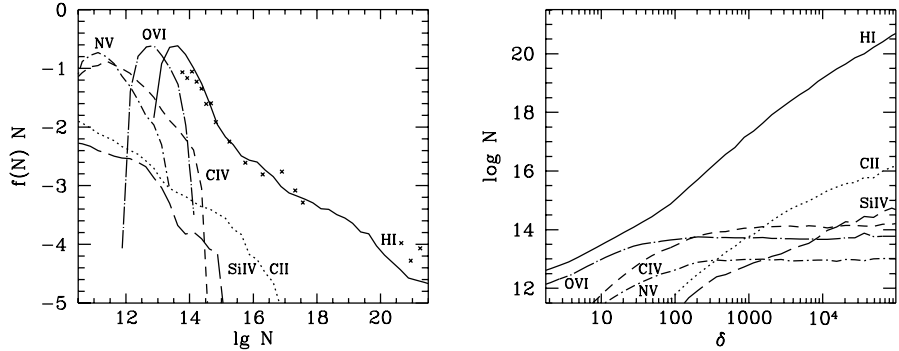,width=13.5cm,angle=0.}
}
\vspace{-4.5cm}
\caption{The left panel shows the fraction of the projected simulation
box  which has column density N  (per ln N). In the right panel 
the mean log N is plotted against the absorption weighted mean
overdensity along the line-of-sight. A homogeneous metallicity of 
0.01 solar is assumed.}
\end{figure}

\section{Column density ratios of different ionic species}

In Figure 3 we show a selection of column density ratios 
as function of HI and CIV column density assuming a homogeneous
metallicity of 0.01 solar. These ratios can be used as diagnostic 
of the UV radiation field and the metallicity of the gas. As
consequence of the tight density-column density correlation the ratio 
of the species  [CII/HI] and [SiIV/HI] drops rather fast
towards lower HI column density. [CIV/HI]  has a pronounced peak 
arround $\log N(HI) \sim$ $14$ to $15$ while [OVI/HI] rises rapidly 
towards small column densities. Recently, there have been measurements
of CII, SiIV, CIV and NV column densities down to $\log N(HI) =
14.5$  by Soingaila \& Cowie (1996). Even so the number of detected
lines are small the measured  [SiIV/CIV] seems to be larger 
than those shown in Figure 3b by a factor of 3 at the relevant CIV column
density while [NV/CIV] is smaller by about the same factor.
A preliminary analysis suggests  that this indicates an ionizing spectrum 
softer than the assumed power law at high energies 
(see Rauch, Haehnelt \& Steinmetz 1996 for a further discussion). 
OVI has not yet been searched  for at small HI column densities. 
If OVI should turn out to be weak below $\log N(HI)=14.5$ 
this would strongly argue for a metallicity gradient towards 
low-density regions.

\begin{figure}[t]
\centerline{
\psfig{file=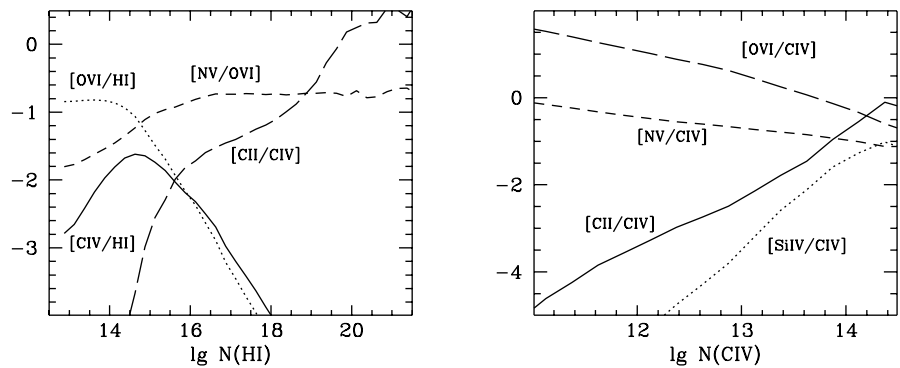,width=14.0cm,angle=0.}
}
\vspace{-4.5cm}
\caption{A set of column density ratios is shown as function of HI 
and CIV column density. A homogeneous metallicity of 
0.01 solar is assumed.}
\end{figure}

%

%
%

\end{document}